\documentclass[pra,twocolumn,showpacs,superscriptaddress]{revtex4}

\begin{document}
\bibliographystyle{apsrev}

\title{Universal simulation of Hamiltonian dynamics for
quantum systems with finite-dimensional state spaces}
\date{\today}
\author{Michael~A.~Nielsen}
\email{nielsen@physics.uq.edu.au}
\affiliation{Centre for Quantum Computer Technology and Department of
Physics, University of Queensland, \\
Brisbane 4072, Queensland, Australia}
\author{Michael~J.~Bremner}
\email{bremner@physics.uq.edu.au}
\affiliation{Centre for Quantum Computer Technology and Department of
Physics, University of Queensland, \\
Brisbane 4072, Queensland, Australia}
\author{Jennifer~L.~Dodd}
\email{jdodd@physics.uq.edu.au}
\affiliation{Centre for Quantum Computer Technology and Department of
Physics, University of Queensland, \\
Brisbane 4072, Queensland, Australia}
\author{Andrew~M.~Childs}
\email{amchilds@mit.edu}
\affiliation{Centre for Quantum Computer Technology and Department of
Physics, University of Queensland, \\
Brisbane 4072, Queensland, Australia}
\affiliation{Center for Theoretical Physics, Massachusetts Institute of 
Technology, Cambridge, MA 02139, USA}  
\author{Christopher~M.~Dawson}
\email{dawson@physics.uq.edu.au}
\affiliation{Centre for Quantum Computer Technology and Department of
Physics, University of Queensland, \\
Brisbane 4072, Queensland, Australia}

\begin{abstract} What interactions are sufficient to simulate
  arbitrary quantum dynamics in a composite quantum system?  Dodd {\em
    et al.} [Phys. Rev. A \textbf{65}, 040301 (R) (2002)] provided a
  partial solution to this problem in the form of an efficient
  algorithm to simulate any desired two-body Hamiltonian evolution
  using any fixed two-body entangling $N$-qubit Hamiltonian, and local
  unitaries.  We extend this result to the case where the component
  systems have $D$ dimensions.  As a consequence we explain how
  universal quantum computation can be performed with any fixed
  two-body entangling $N$-{\em qudit} Hamiltonian, and local
  unitaries.
\end{abstract}


\maketitle

\section{Introduction}
\label{sec:intro}

%
%

A fundamental problem of physics is to determine if there exist
physical systems that are {\em universal}, in the sense that they can
be used to efficiently simulate any other system.  A candidate for
such a universal system was proposed in Deutsch's 1985
paper~\cite{Deutsch85a} in the form of a {\em universal quantum
computer}~\cite{Deutsch85a,Nielsen00a,Preskill98c}.  The purpose of
this paper is to investigate what physical systems are universal for
quantum computation.

%
%

The standard model of a quantum computer consists of $N$ qubits,
prepared in the state $|0\rangle^{\otimes N}$, that can be manipulated
by a sequence of one- and two-qubit operations, and are subsequently
measured in the computational basis.  There are many possible physical
implementations of this model, and in general it is an interesting
problem to determine what critical feature or features of a physical
system enable universal quantum computation.

%
%

In earlier work by Dodd, Nielsen, Bremner and Thew~\cite{Dodd01a} it
was shown that {\em entanglement} is a crucial physical ingredient for
quantum computation.  In particular,~\cite{Dodd01a} showed that the
ability to do local unitary operations together with any {\em fixed}
$N$-qubit two-body entangling Hamiltonian may be used to do universal
quantum computation on those $N$ qubits.

%
%

In this paper we generalize this result to Hamiltonians defined on
{\em qudits}, that is, $D$-dimensional quantum systems spanned by the
states $|0\rangle,\ldots,|D-1\rangle$.  This is of intrinsic interest,
and is also noteworthy because of the much richer structure revealed
in the general proof than in the $D = 2$ case studied
in~\cite{Dodd01a}.

%
%

To state our main result more precisely, we expand an arbitrary
Hamiltonian on $N$ qudits as
\begin{eqnarray} \label{eq:op_expn} H & =
& \sum_{j_1k_1\ldots j_Nk_N} \alpha_{j_1k_1\ldots j_N k_N}
X^{j_1}Z^{k_1} \otimes \ldots \otimes X^{j_N}Z^{k_N}, \nonumber \\
 & & \end{eqnarray} where $j_1,k_1,\ldots,j_N,k_N$ each run from
$0$ through $D-1$, $\alpha_{j_1k_1\ldots j_Nk_N}$ are complex
numbers, and the operators $X$ and $Z$ are $D$-dimensional
generalizations of the familiar Pauli operators, to be defined more
precisely later.  In our work we restrict attention to the case of
Hamiltonians that only include two-body coupling terms, and do
not allow three- or more-body coupling terms.  A {\em two-body coupling}
between a pair of qudits $p$
and $q$ is a term in Eq.~(\ref{eq:op_expn}) of the
form $X^j Z^k \otimes X^{l} Z^{m}$, where neither $X^jZ^k$ nor
$X^{l}Z^{m}$ is equal to the identity, so that the term
acts nontrivially on qudits $p$ and $q$, and acts as the identity
on all other qudits.  In order to generate arbitrary
entanglement in the system it is necessary that each qudit pair
$(s,t)$ must be {\em connected} in the sense that there are coupling
terms in Eq.~(\ref{eq:op_expn}) for each adjacent pair in some
sequence $(s,\ldots,t)$ of qudits in the system.  More explicitly, to
any two-body Hamiltonian one can associate a graph whose vertices
correspond to the qudits in the system, and whose edges connect
vertices representing qudits that are coupled by the Hamiltonian.
A Hamiltonian is said to
be a {\em two-body $N$-qudit entangling Hamiltonian} if the graph is
connected, that is, there is a path between any pair of vertices.  Our
main result is as follows:

\begin{quote} Let $H$ be a given two-body entangling Hamiltonian on
$N$ qudits, and let $K$ be a desired two-body Hamiltonian on $N$
qudits.  Then we have an efficient algorithm to simulate evolution due
to $K$ using only (a) the ability to evolve according to $H$, and (b)
the ability to perform local unitaries (that is, single-qudit
unitaries) on the individual qudits. \end{quote}

%
%

The algorithm we explain below for performing this simulation is only
``efficient'' in the sense of computer science.  That is, it requires
resources polynomial in the number of qudits $N$ in the system.  Our
simulation technique is quite involved, and probably too complicated
to be experimentally practical.  However, the point of principle
demonstrated by our simulation technique is of great importance,
namely, that all two-body $N$-qudit entangling Hamiltonians are
qualitatively equivalent, given the ability to perform local unitary
operations. Thus, in some sense, the ability to entangle can be
regarded as a fundamental physical resource --- a type of ``dynamic
entanglement''\footnote{This term was coined in a 1998 conversation
  between one of us (MAN) and Raymond Laflamme.} --- that can be
utilized to perform interesting processes.  We explore this idea and
make some concrete suggestions for its development in the concluding
section, Section~\ref{sec:discussion}.  Furthermore, our work may
motivate future research on more practically viable methods for doing
universal simulation.

%
%

Antecedents to our work may be identified in many different parts of
the scientific literature.  We now enumerate the different fields in
which antecedents may be found, before giving a detailed account of
the prior work, and how it relates to our own.  The basic techniques
we use are generalizations of standard techniques from nuclear
magnetic resonance (NMR), especially the techniques known as
{\em refocusing} and {\em decoupling}.  The main motivation for our work is
inspired by research into universal gates for quantum computation.
More recently there has been substantial interest within the quantum
information science literature in the problem of determining when
one set of Hamiltonians can be used to simulate another.  This
interest has arisen largely independently of work in the quantum
control literature, where closely related issues are being addressed,
albeit using different techniques and language.

%
%

The main antecedents of our methods are standard NMR techniques for
decoupling and refocusing~\cite{Slichter96a,Ernst94a} that have been
developed and refined over the past half century.  These techniques
have mostly been applied to manipulate specific Hamiltonians, rather
than general classes of Hamiltonians.  Ideas from NMR have been
applied in the quantum computing context by Jones and
Knill~\cite{Jones99a}, and by Leung, Chuang, Yamaguchi and
Yamamoto~\cite{Leung00a}, who considered the problem of efficiently
implementing logic gates using a restricted class of Hamiltonians that
arises naturally in NMR.

%
%

One of the main motivations for our work is the desire to understand
what resources are universal for quantum computation.  Much prior work
has been done on this subject, and many universal sets of gates for
quantum computation are known.  See, for
example~\cite{Barenco95a,Nielsen00a}.  The work most closely related
to ours is independent work of J.-L.~Brylinski and
R.~Brylinski~\cite{Brylinski01a}, who used the representation theory
of compact Lie groups and real algebraic geometry to study the problem
of which two-qudit gates are universal for computation, given the
ability to do single-qudit gates.  In particular, they defined the
class of {\em imprimitive} unitary gates on two qudits to be the gates
that are not of the form $V \otimes W$ or $(V \otimes W)S$, where $V$
and $W$ are single-qudit unitaries, and $S$ is the swap operation.
They showed that any imprimitive gate is universal for quantum
computation, given the ability to also do arbitrary local unitary
operations.  Their results thus imply ours for the case when $N = 2$.
Our results differ from theirs in several ways.  First, even in the
case when $N =2$, the techniques used in our proof are radically
different.  Our techniques are much more elementary, relying only on
basic linear algebra, a simple result from the theory of majorization,
and some very elementary number theory.  Thus, our methods give
different insights into the problem of universality than those
in~\cite{Brylinski01a}.  Second, we consider the case where $N > 2$,
which is potentially of great interest for applications to quantum
computation and quantum control.

%
%

Also related to our results is the work on universal gates by Deutsch,
Barenco and Ekert~\cite{Deutsch95a} and by Lloyd~\cite{Lloyd95a},
where it was shown that almost any two-qubit gate is universal for
quantum computation.  Lloyd sketched a generalization of these results
to the case of qudits, and this sketch has recently been made rigorous
by Weaver~\cite{Weaver00a}.  This work differs from ours in that it
focuses on unitary gates rather than continuous-time Hamiltonian
evolution, and does not result in an explicit characterization of
which sets of unitary gates are universal.  Our work explicitly
determines which two-body Hamiltonians, together with local unitary
operations, are universal.  Furthermore,
in~\cite{Deutsch95a,Lloyd95a,Weaver00a} it was assumed that gates
could be independently applied to {\em any} pair of qudits in the
computer, and thus required the ability to turn on and turn off
interactions between arbitrary pairs of qudits.  By contrast, we
assume only a fixed entangling operation.

%
%

Interest in universal quantum gates has recently motivated interest in
the quantum information science literature in the problem of
simulating one Hamiltonian with another.  Independently of Dodd,
Nielsen, Bremner and Thew~\cite{Dodd01a}, the problem of Hamiltonian
simulation for qubits was considered by D\"ur, Vidal, Cirac, Linden,
and Popescu in~\cite{Dur01a}, where it was shown that all two-qubit
entangling Hamiltonians are qualitatively equivalent, in the sense
that one can be used to simulate the other, given the ability to do
local unitaries.  Wocjan, Janzing and Beth~\cite{Wocjan02a} considered
a {\em specific} Hamiltonian acting on a system containing $N$
spin-$1/2$ particles, and considered the overhead incurred when using
this Hamiltonian to simulate other Hamiltonians.  Bennett, Cirac,
Leifer, Leung, Linden, Popescu and Vidal~\cite{Bennett01a} have
considered the problem of {\em optimal} simulation of one two-qudit
Hamiltonian by another, using general local operations, possibly
including ancillas and measurements.  Thus they considered a different
model than ours, which only involves local unitary operations, and, in
particular, does not require the ability to perform interactions with
local ancilla.  Bennett {\em et al} showed that in the two-qubit case
the two models are, in fact, equivalent.  We also note that the
results in~\cite{Bennett01a} are restricted to the case $N = 2$.
Vidal and Cirac~\cite{Vidal01b} extended the results of Bennett {\em
  et al} by explicitly obtaining the optimal simulation of one
two-qubit Hamiltonian by another in the case where classical
communication between parties is allowed, in addition to the ability
to do local operations, including the use of ancillas and measurement.
They also showed that in the case of two qudits the model where local
unitary operations are allowed is {\em distinct} from the case where
local unitary operations and ancilla are allowed, in the sense that
the latter may be more efficient than the former.  We note that Leung
and Vidal~\cite{Leung01b} have independently obtained results on
problems related to those we consider.  It is worth noting that while
related problems are being addressed in this long list of papers, the
methods used are quite varied, and the different methods may provide
different insights into quantum dynamics.

%
%

Independently of the quantum information science literature there has
been much interest in Hamiltonian simulation in the field of {\em
quantum control}.  A recent overview of work in quantum control may be
found in~\cite{Rabitz00a}.  Of particular interest in this context is
a general set of necessary and sufficient conditions for determining
whether a given set of Hamiltonians can be used to simulate an
arbitrary Hamiltonian (see, for example, Schirmer, Solomon and
Leahy~\cite{Schirmer01a}).  These conditions can be applied to
determine whether, in any specific instance, a collection of
Hamiltonians can be used to simulate an arbitrary Hamiltonian, however
they do not directly speak to the question of what class of
interactions is universal for quantum computation, given the ability
to perform local unitaries.

%
%

Finally, we note that the techniques used in this paper are closely
related to the interesting problem of using a Hamiltonian $H$ to
simulate time-reversed evolution due to the Hamiltonian $-H$.  Results
on this problem have been obtained by Janzing, Wocjan and
Beth~\cite{Janzing01a}, and by Leung~\cite{Leung01a}.

%
%

The structure of our paper is as follows.
Section~\ref{sec:background} introduces background techniques needed
in the main body of the paper, including results on the Pauli group
and majorization.  Section~\ref{sec:two-qudits} explains the $N = 2$
case of the general problem, that is, how any two-qudit entangling
Hamiltonian can be used to simulate any other two-qudit Hamiltonian,
provided local unitaries are allowed.  In
Section~\ref{sec:applications} we explain how this result can be
applied to the general problem of quantum computation on $N$ qudits,
and prove our central result.  Finally, Section~\ref{sec:discussion}
concludes the paper with a summary and discussion of our results, and
a discussion of open problems.

%
%

As the main body of the paper involves a quite extensive construction,
some readers may not wish to wade through all the details.  We have
structured the paper so that such a reader may follow the summaries
provided at the beginning of Sections~\ref{sec:background}
and~\ref{sec:two-qudits}, all of Section~\ref{sec:applications} on
universal quantum computation, and all of the discussion in
Section~\ref{sec:discussion}.

\section{Background}
\label{sec:background}

%
%

We now review the background needed to appreciate the main body of the
paper.  At a first read it may be useful to skip over the proofs, and
pause mainly to appreciate the nomenclature and basic results.
Readers who wish to skip the entire section should note the main
result: given the ability to evolve according to a Hamiltonian $J$ and
perform unitary operations $U_k$ it is possible to simulate evolution
according to a Hamiltonian of the form $\sum_k \alpha_k U_k J
U_k^{\dagger}$, where the $\alpha_k$ are real numbers.  This
composition law for Hamiltonians is the basis for all our later
simulation results.  Note also that throughout the paper we use $=_D$
to indicate equality modulo $D$.  So, for example, $7 =_4 3$, since
$7$ is equal to $3$, modulo $4$.

%
%

The structure of the section is as follows. In
Subsection~\ref{subsec:Pauli} we summarize the relations satisfied by
operators in the Pauli group.
Subsection~\ref{subsec:composition_laws} describes the composition
laws used later in the paper to build up a library of Hamiltonians we
can efficiently simulate given the primitive Hamiltonians initially at
our disposal.  Finally, Subsection~\ref{subsec:majorization} reviews
the basic elements of the theory of majorization, including a
corollary of Uhlmann's theorem crucial to our later analysis.

\subsection{The Pauli group}
\label{subsec:Pauli}

%
%

The $D$-dimensional Pauli group consists of all $D$-dimensional
operators of the form $\omega^l X^j Z^k$, where $j,k,l =
0,\ldots,D-1$, $\omega = \exp(2\pi i/D)$, \begin{eqnarray} X|z\rangle
\equiv |z\oplus 1\rangle; \,\,\,\, Z|z\rangle \equiv \omega^z
|z\rangle, \end{eqnarray} and $\oplus$ denotes addition modulo $D$.
The properties of the Pauli group were investigated in detail by
Gottesman~\cite{Gottesman99b}, and the reader is referred to that
paper for additional information.

%
%

It is worth noting a few simple properties of the Pauli matrices.
First, $X^D = Z^D = I$.  Thus, when writing the Pauli matrices we can
freely interchange expressions like $Z^{D-1}$ and $Z^{-1}$, and
expressions like $Z^{\dagger}$ and $Z^{-1}$.  In a similar vein, note
that $(X^jZ^k)^{\dagger} = Z^{-k}X^{-j}$.  Through most of the paper
we use notation like $Z^{-k}X^{-j}$ in preference to
$(X^jZ^k)^{\dagger}$.

%
%
The basic commutation relations for the Pauli group may be written
\begin{eqnarray} \label{eq:Pauli_comm}
\left( X^j Z^k \right) \left(X^s Z^t \right) = \omega^{ks-jt} 
\left(X^s Z^t \right)\left( X^j Z^k \right).
\end{eqnarray}
We will have very frequent occasion to use these commutation
relations.  In particular, note that $X^jZ^k$ commutes with $X^sZ^t$ if 
and only if $ks =_D jt$.

%
%

Gottesman~\cite{Gottesman99b} studied the properties of the Pauli
group under conjugation by $D$-dimensional unitary operators.  In
particular, he was interested in {\em normalizer operations}, that is,
unitary operations $U$ such that under conjugation by $U$ the Pauli
group is taken to itself.  In Appendix~\ref{app:Gottesman} we
explicitly describe unitary operators performing the following three
conjugation operations:
\begin{eqnarray} X \rightarrow Z, & \,\, & Z \rightarrow X^{-1};
\label{eq:Gott1} \\ X \rightarrow XZ, & \,\, & Z \rightarrow Z;
\label{eq:Gott2} \\ X \rightarrow X^a, & \,\, & Z \rightarrow
Z^{a^{-1}}, \mbox{ when } \gcd(a,D) = 1,  \label{eq:Gott3}
\end{eqnarray} 
where $A \rightarrow B$ means that $UAU^{\dagger} = e^{i \theta} B$
for some phase factor $e^{i \theta}$.  Note that the phase factors are
unimportant for the proof, and will mostly be ignored in the sequel.
These equations imply that the following three conjugation operations
may also be performed:
\begin{eqnarray} X \rightarrow Z^{-1}, & \,\, & Z \rightarrow X;
\label{eq:Gott4} \\ X \rightarrow X, & \,\, & Z \rightarrow XZ;
\label{eq:Gott5} \\ X \rightarrow X^{a^{-1}}, & \,\, & Z \rightarrow
Z^{a}, \mbox{ when }
        \gcd(a,D) = 1.  \label{eq:Gott6}
\end{eqnarray}

%
%

We now use the normalizer operations to prove what we term the {\em
  Pauli-Euclid-Gottesman (PEG) lemma}.  Aside from its interest as
applied in this paper, the PEG lemma is also interesting because it
enables us to explicitly calculate the eigenvalues and eigenvectors of
all elements of the Pauli group, showing a surprising connection
between the Pauli group and Euclid's algorithm (\cite{EuclidBC}, Book
7, Propositions~1 and~2) for finding the greatest common divisor.

%
%

\begin{quote} {\bf Pauli-Euclid-Gottesman Lemma:} For any dimension $D$ 
  and for integers $j$ and $k$ such that $1 \leq j,k \leq D-1$, there
  exists a unitary operator $U$ such that $X^j Z^k \rightarrow
  Z^{\gcd(j,k)}$ under conjugation by $U$.
\end{quote}

%
%

Note, incidentally, that the PEG lemma implies that the eigenvalues of
$X^jZ^k$ are equal to the eigenvalues of $Z^{\gcd(j,k)}$, up to a
global phase which may be calculated from the proof, below.  The
eigenvalues of $Z^{\gcd(j,k)}$ are easily calculable, since $Z$ is
already diagonal.  The eigenvectors of $X^jZ^k$ may also be extracted
from the proof of the PEG lemma, below, where we explain how to
construct the conjugating operation, $U$.

%
%

{\bf Proof:} From Eqns.~(\ref{eq:Gott1})-(\ref{eq:Gott6}) we see that
it is possible to perform the following operations under conjugation
\begin{eqnarray} X^jZ^k & \rightarrow & X^jZ^{k+\alpha j}
\label{eq:Gott7} \\ X^jZ^k & \rightarrow & X^{j+\alpha k} Z^k,
\label{eq:Gott8} \end{eqnarray} where $\alpha$ is any integer.  The
basic idea of the proof is to use these two operations and Euclid's
algorithm on the paired exponents of $X$ and $Z$.  We will give an
example of how this is done, with the general proof following similar
lines.  Consider the operator $X^{104}Z^{80}$.  Recall how Euclid's
algorithm is used to find the greatest common divisor of $104$ and
$80$.  We write $104 = 1 \times 80 + 24$, so $\gcd(104,80) =
\gcd(80,24)$. Next, we write $80 = 3 \times 24 + 8$, so $\gcd(104,80)
= \gcd(24,8)$.  Finally, we write $24 = 3 \times 8$, so $\gcd(104,80)
= 8$.  These steps are easily mimicked with the Pauli operators using
Eqns.~(\ref{eq:Gott7}) and~(\ref{eq:Gott8}).  We have
\begin{eqnarray} X^{104}Z^{80} & \rightarrow & X^{104-1\times
80}Z^{80} = X^{24}Z^{80} \\ X^{24}Z^{80} & \rightarrow & X^{24}Z^{80-3
\times 24} = X^{24} Z^8 \\ X^{24}Z^8 & \rightarrow & X^{24-3\times
8}Z^8 = Z^8.  \end{eqnarray} The general proof proceeds analogously,
using Euclid's algorithm.  {\bf QED}

%
%

A key tool in our analysis is the {\em operator expansion}.  We will
explain in detail how this expansion works for the case of two qudits.
Any operator $J$ on two qudits may
be expanded in the form
\begin{eqnarray} \label{eq:op_expansion}
J = \sum_{j k l m} r_{j k l m} X^j Z^k \otimes
        X^l Z^m,
\end{eqnarray}
where the sum is over the range $j,k,l,m = 0,\ldots,D-1$, and
the coefficients $r_{j k l m}$ may be calculated using
the expression
\begin{eqnarray}
r_{j k l m} =
        \frac{\mbox{tr}\left[
\left( Z^{-k}X^{-j} \otimes Z^{-m}X^{-l} \right)J\right]}
        {D^2}.
\end{eqnarray}
In general it is useful to introduce the convention that the indices in
sums always range over $0,\ldots,D-1$, unless otherwise noted.

Eq.~(\ref{eq:op_expansion}) applies for any operator, however,
Hermitian operators satisfy additional constraints on the form of the
coefficients $r_{j k l m}$.  For example, if a term of the form
$\alpha Z^k \otimes Z^m$ appears in the operator expansion, then its
Hermitian conjugate $\alpha^* Z^{-k} \otimes Z^{-m}$ must also appear
in the operator expansion.  In general, terms in the operator
expansion of a Hermitian operator appear in Hermitian conjugate pairs.

%
%

The operator expansion may be used to establish a useful identity 
satisfied by any operator $J$ on a single qudit,
\begin{eqnarray} \label{eq:Schur}
\sum_{jk} \left( X^j Z^k \right) 
        J \left( Z^{-k} X^{-j} \right) = D \, \mbox{tr}(J) I.
\end{eqnarray}
The identity Eq.~(\ref{eq:Schur})
is well-known in quantum information science from the
properties of the depolarizing channel for $D$-dimensional systems.
The identity may be verified by direct calculation, or by substituting
an operator expansion for $J$.  Eq.~(\ref{eq:Schur}) may also be extended to 
multiple qudits.  For our purposes all that matters is the two-qudit case, 
which reads:
\begin{eqnarray} \label{eq:Schur2}
& & \sum_{jklm} \left( X^j Z^k \otimes X^l Z^m \right)
        J\left( Z^{-k} X^{-j} \otimes Z^{-m} X^{-l} \right)
        \nonumber \\
& & = D^2 \mbox{tr}(J) I \otimes I.
\end{eqnarray}

%
%

We conclude this subsection with a brief digression, noting that a
beautiful alternate proof of Eq.~(\ref{eq:Schur}) may be obtained by
applying Schur's lemma from group representation
theory~\cite{Serre77a}.  Let $G_D$ denote the Pauli group in $D$
dimensions, and note that
\begin{eqnarray}
\sum_{jk} \left( X^j Z^k \right) J 
        \left( Z^{-k} X^{-j} \right) = \frac{1}{D} 
        \sum_{U \in G_D} U J U^{\dagger}.  \end{eqnarray} The factor
$1/D$ on the right-hand side arises because of the phases $\omega^l$
in front of a general member of the Pauli group, $\omega^l X^j Z^k$.
The right-hand side of this equation commutes with any $U \in G_D$.
The result follows from Schur's lemma if we can prove that $G_D$ is
irreducible.  Suppose $G_D$ is reducible, so that there exists a
non-trivial subspace of the qudit state space stable under the
operations in $G_D$.  Let $P$ denote the projector onto that subspace.
Because the subspace is stable it follows that $Z P Z^{-1} = P$, and
thus $Z$ commutes with $P$.  It follows that $P$ can be diagonalized
in the basis $|0\rangle,\ldots,|D-1\rangle$, and thus the stable
subspace is spanned by a strict subset of
$|0\rangle,\ldots,|D-1\rangle$.  Suppose $|z\rangle$ is in the stable
subspace, but $|z\oplus 1\rangle$ is not.  But $X|z\rangle = |z\oplus
1\rangle$, so the subspace is not stable, which is a contradiction.
This completes the proof of Eq.~(\ref{eq:Schur}).

\subsection{Composition laws for Hamiltonian simulation}
\label{subsec:composition_laws}

%
%

The basic idea employed in the main part of the paper is to use our
primitive set of operations and a small number of {\em composition
laws} to build up a library of Hamiltonian evolutions we can simulate.
We now explain the composition laws that we use, adapting
from~\cite{Dodd01a}.

%
%

(A) Imagine we can evolve according to the Hamiltonian $J$, and
perform unitary operations $U$ and $U^{\dagger}$.  Then it follows
from the identity $e^{-itUJU^{\dagger}} = U e^{-itJ}U^{\dagger}$ that
we can exactly simulate evolution according to the Hamiltonian
$UJU^{\dagger}$.

%
%

(B) Imagine we can evolve according to the Hamiltonians $J_1$ and
$J_2$.  Then we can simulate evolution due to $J_1 + J_2$ for small
times $\Delta$, due to the approximate identity
\begin{eqnarray} \label{eq:approx_identity}
e^{-i\Delta(J_1+J_2)} \approx e^{-i \Delta J_1}e^{-i\Delta J_2}.
\end{eqnarray}
We treat this identity as though it is exact.  This is justified, since to
simulate a Hamiltonian for a time $t$ it suffices to perform $n$
separate simulations of a time $\Delta \equiv t/n$ each, giving an error of
$n \times O(\Delta^2) = O(t\Delta)$.  This error
can thus be made arbitrarily small
by making $\Delta$ sufficiently small.  Further remarks on the error
analysis are made for the qubit case in~\cite{Dodd01a}, and these results
carry over directly to the qudit case.

%
%

(C) Imagine we can evolve according to the Hamiltonian $J$.  Then, by
appropriate timing, we can exactly simulate evolution according to
$\lambda J$ for any $\lambda > 0$. 

%
%

(D) Imagine we can evolve according to the Hamiltonian $J$.  Then we
can evolve according to the Hamiltonian $-J$.  We will explicitly prove this
only for the case of two-qudit Hamiltonians $J$, however the proof easily
generalizes.  Note that we can rewrite Eq.~(\ref{eq:Schur2}),
\begin{eqnarray}
-J & = & \sum \left( X^j Z^k \otimes X^l Z^m \right)
        J \left( Z^{-k} X^{-j} \otimes Z^{-m} X^{-l} \right)
        \nonumber \\
& & -D^2 \mbox{tr}(J) I \otimes I,
\end{eqnarray}
where we have extracted the $-J$ by taking the sum on the right-hand side 
over all terms except $(j,k,l,m) = (0,0,0,0)$. 
Physically, the term $-D^2 \mbox{tr}(J) I \otimes I$ 
is an unimportant rescaling
of the energy and can be neglected.  The other terms in the expansion
for $-J$ are all easily simulated using techniques (A) and (B) above.  Note
that the complexity of the simulation scales as $O(D^4)$.

%
%

The above observations (A)--(D) may be summarized in a single equation
as follows. Given the ability to perform evolution according to the
Hamiltonian $J$ and the ability to perform unitaries $U_k$, it is
possible to simulate evolution according to a Hamiltonian of the form
\begin{eqnarray} \label{eq:simulation}
\sum_k \alpha_k U_k J U_k^{\dagger},
\end{eqnarray}
where the $\alpha_k$ can be arbitrary real numbers.

\subsection{Majorization}
\label{subsec:majorization}

%
%

The final area of background we shall need is the theory of {\em
majorization}, whose basic elements we now review,
following~\cite{Nielsen01a}.  More detailed introductions to
majorization may be found in Chapters~2 and~3 of \cite{Bhatia97a},
\cite{Marshall79a}, and \cite{Alberti82a}.  
Suppose $x = (x_1,\ldots,x_D)$ and $y =
(y_1,\ldots,y_D)$ are two $D$-dimensional real vectors.  The relation
``$x$ is majorized by $y$'' is intended to capture the notion that $x$
is more mixed (i.e. disordered) than $y$.  To make the formal
definition we introduce the notation $\downarrow$ to denote the
components of a vector rearranged into non-increasing order, so
$x^{\downarrow} = (x^{\downarrow}_1,\ldots,x^{\downarrow}_D)$, where
$x^{\downarrow}_1 \geq x^{\downarrow}_2 \geq \ldots \geq
x^{\downarrow}_D$.  We say that $x$ is majorized by $y$, and write $x
\prec y$, if
\begin{eqnarray}
\sum_{j=1}^k x^{\downarrow}_j \leq \sum_{j=1}^k y^{\downarrow}_j,
\end{eqnarray}
for $k = 1,\ldots,D-1$, and with the inequality holding with
equality when $k = D$.

%
%

The notion of majorization can be extended in a natural way to
Hermitian operators.  We say that the Hermitian operator $A$ is
majorized by the Hermitian operator $B$, and write $A \prec B$, if the
spectrum $\lambda(A)$ of eigenvalues of $A$ is majorized by the
spectrum $\lambda(B)$ of eigenvalues of $B$, where we regard the
spectra $\lambda(A)$ and $\lambda(B)$ as vectors.  So, for example,
\begin{eqnarray}
\left[ 
\begin{array}{cc} \frac{1}{2} & 0 \\ 0 & \frac{1}{2} \end{array}
\right]
& \prec &
\left[ \begin{array}{cc}
        \frac{1}{2} & \frac{1}{2} \\
        \frac{1}{2} & \frac{1}{2}
        \end{array}
\right],
\end{eqnarray}
since the spectra of the two matrices satisfy the majorization criterion,
$(1/2,1/2) \prec (1,0)$.

%
%

It is not immediately obvious how this definition of operator
majorization connects to any natural notion of comparative disorder.
There is a beautiful result due to Uhlmann~\cite{Uhlmann71a} (see also
the reviews~\cite{Wehrl78a} and~\cite{Nielsen01b}) that provides such
a connection.  Uhlmann's theorem states that $A \prec B$ if and only
if $A = \sum_n p_n U_n B U_n^{\dagger}$, where the $U_n$ are unitary
operators, and the $p_n$ form a probability distribution.  That is, $A
\prec B$ if and only if $A$ can be obtained from $B$ by mixing
together operators unitarily equivalent to $B$.  Two important points
about the proof of Uhlmann's theorem are that the procedure for
finding the $p_n$ and $U_n$ is {\em constructive}, and, furthermore,
there are at most $D^2$ operators $U_n$.

%
%

We now observe that Uhlmann's theorem has a beautiful corollary when
applied to {\em any} two traceless Hermitian operators $A$ and 
$B$:\footnote{Our thanks go to Ben Schumacher, who contributed to the
discovery of this theorem.}
\begin{quote} {\bf Theorem:} Let $A$ and $B$ be any two traceless
Hermitian operators, and assume $B \neq 0$.
Then $A \prec c B$ for some positive constant $c$.  Uhlmann's theorem
then gives an algorithm to find a set of at
most $D^2$ unitary operators $U_n$, and $c_n > 0$, such that
\begin{eqnarray}
A = \sum_n c_n U_n B U_n^{\dagger}.
\end{eqnarray}
\end{quote}

%
%

As an example of this theorem in action, consider that for any $(j,k)
\neq (0,0)$ and $(l,m) \neq (0,0)$ there exist $U_n$ and $c_n$ such
that:
\begin{eqnarray}
& & X^j Z^k + Z^{-k} X^{-j} \nonumber \\
& = & \sum_n c_n U_n \left( X^l Z^m + Z^{-m} X^{-l} \right) U_n^{\dagger}.
\end{eqnarray}
Using the techniques of the previous subsection, notably
Eq.~(\ref{eq:simulation}), we see that this equation can be
interpreted as providing a means of simulating the Hamiltonian $X^j
Z^k + Z^{-k} X^{-j}$ given the Hamiltonian $X^l Z^m + Z^{-m}X^{-l}$
and the ability to perform the unitary operations $U_n$.  On its own,
this is not an especially useful simulation result!  However, similar
but more sophisticated variants on this idea will be used in our later
construction.

%
%

{\bf Proof:} 
The case when $A = 0$ follows by noting that $0 \prec cB$ for all $c > 0$,
so we assume $A \neq 0$.  
We aim to show that
$\lambda(A) \prec c \lambda(B)$, for some $c > 0$.  Choose
\begin{eqnarray} c \equiv \max_{k=1,\ldots,D-1} \frac{\sum_{j=1}^k
\lambda^{\downarrow}_j(A)}
        {\sum_{j=1}^k \lambda^{\downarrow}_j(B)}.
\end{eqnarray}
Since $A$ and $B$ are traceless and not equal to $0$, it follows that $c > 0$.
For $k = 1,\ldots,D-1$ we have
\begin{eqnarray}
\sum_{j=1}^k \lambda^{\downarrow}_j(A) & = &
        \frac{\sum_{j=1}^k \lambda^\downarrow_j(A)}
        {\sum_{j=1}^k \lambda^{\downarrow}_j(B)}
        \sum_{j=1}^k \lambda^{\downarrow}_j(B) \\
& \leq & c \sum_{j=1}^k \lambda^{\downarrow}_j(B).
\end{eqnarray}
Finally, note that
\begin{eqnarray}
\sum_{j=1}^D \lambda^{\downarrow}_j(A) = 0 =
        \sum_{j=1}^D c \lambda^{\downarrow}_j(B),
\end{eqnarray}
which completes the proof.  {\bf QED}

\section{Two-qudit Hamiltonian simulation}
\label{sec:two-qudits}

%
%

In this section we study universal simulation with two-qudit
Hamiltonians, that is, Hamiltonians of the form
\begin{eqnarray} \label{eq:Hamiltonian}
H = \sum_{jklm} \alpha_{jk lm} X^j Z^k \otimes X^l Z^m.
\end{eqnarray}
We show that provided this Hamiltonian has a non-zero coupling term, 
that is, a term not of the form $I \otimes (\cdot)$ or 
$(\cdot) \otimes I$, then $H$ and local unitary operations 
can be used to simulate {\em any} other
two-qudit Hamiltonian $K$.

%
%

The basic idea of the proof is to use the composition laws of
Subsection~\ref{subsec:composition_laws} to increase the library of
Hamiltonians that can be simulated.  It will be convenient to use the
notation $H_1,H_2,\ldots$ to denote the different Hamiltonians that we
show how to simulate.  The construction is rather complicated, for
which reason we break it up into steps.  This separation into steps
makes it convenient to introduce some global notational conventions.
Terms like $j,k,l,m,n,r,s,t$ are specific to each step, and sometimes
to individual lines in the proof, often being used as dummy variables,
with the meaning to be determined from context.  Terms like
$a,b,c,d,f$ carry over from one step to another.  All of these terms
($j,k,\ldots$ and $a,b,\ldots$) are consistently integers in the range
$0,\ldots,D-1$.

%
%

The general strategy through most of the proof is to gradually eliminate
more and more terms from the Hamiltonian, while keeping particular
desired couplings.  At the end of the proof we are able to simulate a
Hamiltonian of an especially simple form, which can then be used to
build up arbitrarily complicated Hamiltonians.
We now give an outline of the proof.  Note that the numbering scheme
used in the outline is mirrored in the numbering scheme used in the detailed
explanation of the proof given below in 
Subsection~\ref{subsec:detailed_proof}.
\begin{enumerate}

\item We show that $H$ and local unitaries can be used to simulate a
Hamiltonian $H_1$ that contains a $Z^a \otimes Z^b$ coupling term.
This term is the focus of most of the remaining steps of the proof, as
we try to eliminate most of the other coupling terms from the
Hamiltonian.

\item We show that $H_1$ and local unitaries can be used to eliminate
terms in $H_1$ not of the form $Z^j \otimes Z^k$, and thus to simulate
a Hamiltonian $H_2$ of the form $\sum_{jk} \alpha_{jk} Z^j \otimes
Z^k$, which still contains the non-zero coupling
$\alpha_{ab} Z^a \otimes Z^b$.

\item We show that $H_2$ and local unitaries can be used to simulate a
Hamiltonian of the form $H_3 = \sum_n \beta_n (Z^c \otimes Z^d)^n$,
with at least one non-zero coupling coefficient $\beta_f$.

\item We show that $H_3$ and local unitaries can be used to simulate a
Hamiltonian of the form $H_6 = \kappa Z^a \otimes Z^b + \kappa^*
Z^{-a} \otimes Z^{-b}$, for any complex $\kappa$.

\item We show that $H_6$ and local unitaries can be used to simulate $H_8
= (Z^a + Z^{-a}) \otimes (Z^b + Z^{-b})$.

\item Using the corollary to Uhlmann's theorem we show that $H_8$
can be used to simulate any Hamiltonian of the form $J \otimes J'$,
where $J$ and $J'$ are arbitrary traceless Hermitian operators.
Any two-qudit Hamiltonian can be expressed as a sum of terms of
this form, together with local interactions, so we conclude that 
any two-qudit Hamiltonian can be simulated using
$H$ and local unitaries.

\end{enumerate}

%
%

This construction is complex, and a detailed efficiency analysis is
not especially enlightening.  Nonetheless, from the proof below it
follows that the total simulation requires a number of periods of
evolution due to $H$ which is {\em polynomial} in the dimension $D$.
This can be seen by examining each step in the construction and
verifying that they involve only a summation $\sum_k \alpha_k (U_k
\otimes V_k) J (U_k \otimes V_k)^{\dagger}$ over at most polynomially
many terms in $D$, where the $U_k$ and $V_k$ are local unitaries, $J$
is some entangling Hamiltonian that we are already able to simulate,
and the coefficients $\alpha_k$ are also polynomial in $D$.

%
%

It is worth noting that the proof can be substantially simplified if
one assumes that $D$ is prime.  The reason this simplification occurs
is because in prime dimensions all non-trivial elements of the Pauli
group are equivalent to one another by unitary conjugation.  Thus,
given a non-zero coupling term in the Hamiltonian it is easy to
simulate a Hamiltonian in which a non-zero coupling of the form $Z
\otimes Z$ appears.  Given this, steps $1$ and $3$ can be
considerably simplified.  We describe in detail how this
simplification occurs in Subsection~\ref{subsec:prime}.

\subsection{Detailed proof}
\label{subsec:detailed_proof}

\subsubsection{Simulating a Hamiltonian with a non-zero $Z^a \otimes Z^b$
coupling}

%
%

By assumption, our Hamiltonian includes a coupling term of the form
$X^jZ^k \otimes X^lZ^m$ with a non-zero coefficient.  If $j \neq 0$ or
$l \neq 0$ then the PEG lemma and Eq.~(\ref{eq:Gott1}) imply that by
performing local unitary conjugation we can convert this to a coupling
term of the form $Z^a \otimes Z^b$ with a non-zero coefficient.  Let
$H_1$ be the Hamiltonian that results when this conjugation is
performed.  It will be convenient for our later discussion to fix
coprime $c$ and $d$ such that $c/d = a/b$, that is, $c/d$ is $a/b$ in
lowest common terms.  It will also be convenient to define $f$ such
that $a = fc$ and $d = fb$.

\subsubsection{Simulating a Hamiltonian of the form $\sum_{jk} 
\alpha_{jk} Z^j \otimes Z^k$}

We have shown that $H_1$ contains a coupling term of the form $Z^a
\otimes Z^b$, however it could also contain many other coupling terms.
We aim to eliminate these other terms, while keeping the coupling $Z^a
\otimes Z^b$.  In particular, we now explain how to eliminate those
terms containing $X$ or a power of $X$.  Note that given the ability
to do $H_1$ we can simulate
\begin{eqnarray}
H_2 = \sum_{lm} (Z^l \otimes Z^m) H_1 (Z^{-l} \otimes Z^{-m}).
\end{eqnarray}
To evaluate this sum we first use the commutation relations for 
the generalized
Pauli operators and then the observation that 
$\sum_{t} \omega^{st} = D$ when $s =_D 0$, and $\sum_t \omega^{st} = 0$
otherwise.  Using these facts we see that $H_2$ has the form
\begin{eqnarray}
H_2 = \sum_{jk} \alpha_{jk} Z^j \otimes Z^k.
\end{eqnarray}
The term $Z^a \otimes Z^b$ in $H_1$ was non-zero, so $\alpha_{ab} \neq 0$.

\subsubsection{Elimination of all terms not of the form $(Z^c \otimes Z^d)^n$}

%
%

The next step of the proof is to eliminate all the terms in $H_2$
which are not powers of $(Z^c \otimes Z^d)$.  Note that we know there
is at least one non-zero term of this form, the term $Z^a \otimes Z^b
= (Z^c \otimes Z^d)^f$.  The key to this is a simple number-theoretic
lemma:

%
%

\begin{quote} {\bf Lemma:} Suppose $\gcd(l,m) = 1$.  Then 
$j m =_D   k l$ if and only if there exists $n$ such that
\begin{eqnarray}
j =_D nl, \mbox{ and } \,\,\,\,  k =_D nm.
\end{eqnarray}
\end{quote}

We will give two proofs of this lemma.  The first is a constructive
proof that only involves elementary number theory.  The second proof
is in some sense more elegant, in that it invokes the PEG lemma, and
makes use of notions of linear algebra.  The second proof is given in
Appendix~\ref{app:number-theory}.

{\bf Proof:} The reverse implication follows by a simple
substitution, so we prove only 
the forward implication.
Since $\gcd(l,m) = 1$ there exist integers $r$ and $s$
such that $rl+sm = 1$.  Now choose $n \equiv jr + ks$.
Then we have
\begin{eqnarray}
nl & =_D & j r l + k s l\\
   & =_D & j r l + j s m \\
   & =_D & j(rl+sm) \\
   & =_D & j,
\end{eqnarray}
as required.  A similar calculation shows that $nm =_D k$.
{\bf QED}

%
%

Applying our composition laws we see that we can simulate
\begin{eqnarray}
H_3 = \sum_{l} (X^{-d} \otimes X^c)^l  H_2 
(X^d \otimes X^{-c})^l.
\end{eqnarray}
Applying the commutation relations for the Pauli matrices this simplifies to
\begin{eqnarray}
H_3 = \sum_{jk} \alpha_{jk} \left[ \sum_{l} \omega^{(dj-ck)l} \right]
        Z^j \otimes Z^k.
\end{eqnarray}
Note that the sum over $l$ is zero unless $dj =_D ck$.  By the lemma, this is
the case if and only if $j =_D nc$ and $k =_D nd$ for some $n$, and thus
$Z^j \otimes Z^k = (Z^c \otimes Z^d)^n$ for some $n$.
Thus $H_3$ has the form
\begin{eqnarray}
H_3 = \sum_n \beta_n (Z^c \otimes Z^d)^n,
\end{eqnarray}
where the $\beta_n$ are complex numbers.  Recall that $a
= fc$ and $b = fd$, so $\beta_f \propto \alpha_{ab}$, and 
there is at least one non-zero coupling term in $H_3$.

\subsubsection{Simplifying to a sum of at most two terms}

%
%

Our next task is to eliminate nearly all the coupling terms in $H_3$.
First, we set up some notation.  Since $\gcd(c,d) = 1$ we can
choose $l$ and $m$ such that $lc + md
= 1$.  It will be convenient to write the coefficients $\beta_n$ as a
$D$-dimensional vector, that is $\vec \beta =
(\beta_0,\beta_1,\ldots,\beta_{D-1})$, where we use the convention
that expressions like $(x,y,z,\ldots)$ denote column vectors.  It will
also be convenient to use the notation $\vec e_0,\ldots,\vec e_{D-1}$
for the unit vectors in this $D$-dimensional vector space, and to
identify $\vec e_{-j}$ with $\vec e_{D-j}$.  So, for example, $\vec
e_1 = (0,1,0,0,\ldots,0)$, and $\vec e_{-2} = \vec e_{D-2}$.  Note
that the constraint that $H_3$ is Hermitian implies that $\vec \beta^*
= P \vec \beta$, where $P \vec e_n = \vec e_{-n}$ for
$n=0,\ldots,D-1$.

%
%

Next, suppose $\vec \gamma = (\gamma_0,\ldots,\gamma_{D-1})$ is a real
vector.  Using our composition laws we can simulate the Hamiltonian
\begin{eqnarray} \label{eq:h4}
H_4 = \sum_{j} \gamma_j (X^{-l} \otimes X^{-m})^j H_3
        (X^{l} \otimes X^{m})^j.
\end{eqnarray}
Our strategy will be to choose $\vec \gamma$ in such a way that
$H_4$ has an especially simple form.  Applying the commutation
relations for the Pauli operators gives
\begin{eqnarray}
H_4 & = & \sum_{jn} \gamma_j \beta_n \omega^{(lc+md)jn} (Z^c
        \otimes Z^d)^n \\
 & = & \sum_{n} \delta_n (Z^c \otimes Z^d)^n,
\end{eqnarray}
where in the second step we used the fact that $lc+md=1$, and we
define
\begin{eqnarray}
\delta_n \equiv \beta_n \sum_j \omega^{nj} \gamma_j.
\end{eqnarray}
The sum on the right-hand side is most conveniently written in matrix
form as $M \vec \gamma$, where $M$ is the matrix with
entries $M_{nj} \equiv \omega^{nj}$.  Up to a constant $M$ is just
the matrix representation of the discrete Fourier transform, which is
easily inverted, so we can
choose $\vec \gamma$ such that $M \vec \gamma = \vec e_f + \vec e_{-f}$.

%
%

Recall that the $\gamma_j$ in Eq.~(\ref{eq:h4}) must be real in order for
a simulation of $H_4$ to be possible.  We now use a symmetry argument to
show that this is the case.
Note that $M \vec \gamma = \vec e_f + \vec e_{-f}$, by
definition of $\vec \gamma$.  Since $P$, $\vec e_f$ and $\vec e_{-f}$ 
are real, 
\begin{eqnarray}
M \vec \gamma = 
\vec e_f + \vec e_{-f} = \left( P(\vec e_f+\vec e_{-f})\right)^*
\end{eqnarray}
Next, from $M\vec \gamma = \vec e_f+\vec e_{-f}$ and $P^* = P$ we
obtain
\begin{eqnarray}
\left( P(\vec e_f+\vec e_{-f})\right)^* = PM^* \vec \gamma^*.
\end{eqnarray}
Combining these results we see that $PM^* \vec \gamma^* = M \vec \gamma$.
Observing that $PM^* = M$ we obtain $M \vec \gamma^* = M \vec \gamma$,
and thus $\vec \gamma$ is real.

%
%

Summarizing, we have obtained the ability to simulate a Hamiltonian
\begin{eqnarray}
H_4 = \beta Z^a \otimes Z^b + \beta^*
        Z^{-a} \otimes Z^{-b},
\end{eqnarray}
where $\beta \equiv \beta_f$, and the fact that $\beta_{-f} = \beta^*$
follows from the fact that $H_4$ is Hermitian.  Conjugating by $X^{-1}
\otimes I$ we also obtain the ability to simulate the Hamiltonian
\begin{eqnarray}
H_5 = \beta \omega^a Z^a \otimes Z^b + 
        \left(\beta \omega^a\right)^*
        Z^{-a} \otimes Z^{-b}.
\end{eqnarray}
However, note that {\em any} complex number $\kappa$ can be formed from
real linear combinations of $\beta$ and $\beta \omega^a$, so by taking
appropriate real linear combinations of $H_4$ and $H_5$ we see that
we can simulate any Hamiltonian of the form
\begin{eqnarray}
H_6 = \kappa Z^a \otimes Z^b + \kappa^*
        Z^{-a} \otimes Z^{-b}.
\end{eqnarray}

\subsubsection{Simulation of a tensor product Hamiltonian}

%
%

Applying Eq.~(\ref{eq:Gott3}) to the second qudit we see that we
can simulate any Hamiltonian of the form
\begin{eqnarray}
H_{7 \pm} = \kappa Z^{a} \otimes Z^{\pm b} + \kappa^*
        Z^{-a} \otimes Z^{\mp b},
\end{eqnarray}
and it follows by taking linear combinations that we can simulate
\begin{eqnarray}
H_8 = (Z^a+Z^{-a}) \otimes (Z^b +Z^{-b}).
\end{eqnarray}

\subsubsection{Simulation of {\em any} Hamiltonian}

%
%

Note that $Z^a + Z^{-a}$ and $Z^b + Z^{-b}$ are non-zero, traceless,
Hermitian operators, so by the corollary to Uhlmann's theorem we can
simulate {\em any} Hamiltonian of the form $J \otimes J'$, where $J$
and $J'$ are arbitrary traceless, Hermitian operators.  The operator
expansion implies that an arbitrary two-qudit Hamiltonian can be formed
as a real linear combination of such Hamiltonians, together with
single-qudit terms of the form $J \otimes I$ or $I \otimes J'$.  Thus,
with the ability to perform $H$ and local unitary operations we can
simulate an arbitrary two-qudit Hamiltonian.

\subsection{The case where $D$ is prime}
\label{subsec:prime}

%
%

The proof just given can be substantially simplified in the case where
$D$ is prime.  We now sketch how the simplified proof goes.  The
reason for the simplification is that any non-trivial element $X^j
Z^k$ of the Pauli group is equivalent under conjugation to $Z$.  To
see this, note that if $j \neq 0$ and $k \neq 0$ then, using the PEG
lemma it is possible to conjugate $X^j Z^k$ to $Z^l$, up to a phase
factor, for some $l$ such that $1 \leq l \leq D-1$.  Similarly, if $k
= 0$ then we can conjugate to some such $Z^l$ using~(\ref{eq:Gott1}),
while if $j = 0$ then the term is already in this form.  In turn this
may be conjugated to $Z$ using Eq.~(\ref{eq:Gott3}), since $l$ is
co-prime to $D$ when $D$ is prime.  It follows that in step~1 of the
above proof we can show that it is possible to simulate a Hamiltonian
$H_1$ that contains a $Z \otimes Z$ coupling term.  Step~2 proceeds
exactly as before, and results in a Hamiltonian of the form $H_2 =
\sum_{jk} \alpha_{jk} Z^j \otimes Z^k$, such that $\alpha_{11} \neq
0$.

%
%

Step~3 of the preceding proof is substantially simplified.  In
particular, we note that it is possible to simulate the Hamiltonian
\begin{eqnarray}
H_3 & = &
    \sum_{l} (X^l \otimes X^{-l}) H_2  (X^{-l} \otimes X^l)
 \\
 & = & \sum_{jkl} \alpha_{jk} \omega^{l(k-j)} Z^j \otimes Z^k \\
 & = & D \sum_j \alpha_{jj} (Z \otimes Z)^j.
\end{eqnarray}
The remainder of the proof can then be completed as before.

\section{Applications to universal quantum computation}
\label{sec:applications}

%
%

We have shown that any two-qudit entangling Hamiltonian, together with
local unitary operations, may be used to simulate any other two-qudit
Hamiltonian.  We now extend this result to the problem of universal
quantum computation on $N$ qudits.  In particular, we show that any
two-body $N$-qudit entangling Hamiltonian, together with local
unitaries, can be used to perform universal quantum computation.

%
%

The basic strategy follows the method presented in~\cite{Dodd01a}.
The idea is to reduce the problem to the two-qudit case already
solved.  To do this, we divide the system into a {\em principal
system} $P$ consisting of two qudits coupled by the Hamiltonian $H$ of
the entire system, and the {\em remainder} of the system, denoted $S$.
Our techniques generalize the results in~\cite{Jones99a,Leung00a},
which are themselves generalizations of standard techniques from NMR.
The basic idea is to turn off all the interactions between $P$ and
$S$, and within $S$, leaving only the interactions present in $P$.  We
will refer to such a suppression of interactions as {\em decoupling}.
The remaining interactions can then be used, as before, to simulate
arbitrary dynamics on the two qudits in $P$.  Thus it is possible to
simulate arbitrary dynamics on {\em any} two qudits coupled by $H$.
Finally, an arbitrary interaction between qudits $s$ and $t$ may be
effected by performing a sequence of {\sc swap} gates between the
qudits connecting $s$ and $t$ (in the sense defined in
Section~\ref{sec:intro}), applying the desired interaction, and
swapping back.  Note that such a sequence of {\sc swap} operations can
be performed using the method already described for simulating quantum
gates.  In this way we can effect any two-qudit Hamiltonian between
any pair of qudits in the system, and thus perform universal quantum
computation.

%
%

The obvious technique for achieving decoupling is to eliminate
couplings between $P$ and $S$, and within $S$, one at a time, using
techniques along the lines of those used to simulate one two-qudit
Hamiltonian with another.  Unfortunately, this procedure is not
efficient, for reasons we now explain.  As an example, suppose $P$
consists of two qudits, labeled $A$ and $B$, and $S$ consists of two
qudits, $E$ and $F$.  Then interactions between qudit $E$ and the
remainder of the system may be effectively turned off by simulating
the Hamiltonian \begin{eqnarray} H' = \frac{\sum_{U} U_E H
    U_E^{\dagger}}{D^2}, \end{eqnarray} where $U$ runs over all Pauli
matrices $X^jZ^k$ and the $E$ indicates that $U$ is being applied to
the qudit $E$.  While we can in principle turn off all interactions
in this way, the resulting procedure is not efficient.  To see this,
notice that turning off all the couplings to the qudit $E$ required a
sum over $D^2$ terms, each a conjugated form of the Hamiltonian $H$.
For an $N$-qudit system generalizing this procedure in the obvious way
would require a sum over $D^N$ terms.  The corresponding simulation
would therefore have exponential complexity, which is not efficient.

%
%

Fortunately, much more efficient techniques for decoupling can be
devised.  In this section we explain two such techniques.
Subsection~\ref{subsec:arbitrary} explains how the decoupling can be
performed for a completely {\em arbitrary} two-qudit Hamiltonian,
while Subsection~\ref{subsec:local} explains how the procedure for
decoupling can be substantially simplified and made more efficient
when the Hamiltonian has the localized structure
found in most physical systems.

\subsection{The case of arbitrary two-qudit interactions}
\label{subsec:arbitrary}

%
%

Suppose $H$ is an arbitrary two-qudit entangling Hamiltonian.  We now
explain how to efficiently eliminate all couplings between a
principal system $P$ and the remainder of the system $S$, and to
eliminate all couplings internal to $S$, while leaving the couplings
within $P$ invariant.  The method is a straightforward generalization
of that described for qubits by Dodd {\em et al}~\cite{Dodd01a}.  Let
$U$ run over all Pauli matrices $X^j Z^k$.  Define $U_{S}$ to be the
tensor product of identical operators $U$ acting ditwise on the qudits
in $S$.  We form the Hamiltonian
\begin{eqnarray}
H' = \frac{1}{D^2} \sum_U U_{S} H U_{S}^{\dagger},
\end{eqnarray}
and observe that $H'$
leaves the Hamiltonian on $P$ invariant, but eliminates all coupling
terms between $P$ and $S$, and all single-qudit terms acting 
within $S$.

%
%

We now explain a recursive construction to eliminate all remaining
couplings in $S$.  First, we break the block $S$ into two blocks $S_0$
and $S_1$ of approximately equal size.  We decouple $S_0$ and $S_1$ by
forming the Hamiltonian
\begin{eqnarray}
H^{''} = \frac{1}{D^2} \sum_U U_{S_0} H^{'} U_{S_0}^{\dagger}.
\end{eqnarray}
Next, we break $S_0$ into two blocks $S_{00}$ and $S_{01}$ of
approximately equal size, and break $S_1$ into two blocks
$S_{10}$ and $S_{11}$ of approximately equal size.  We can
decouple $S_{00}$ from $S_{10}$, and $S_{01}$ from $S_{11}$
in a single step by forming the Hamiltonian
\begin{eqnarray}
H^{'''} = \frac{1}{D^2} \sum_U \left( U_{S_{00}} \otimes U_{S_{10}} 
        \right) H^{''} \left( U_{S_{00}} \otimes U_{S_{10}}\right)^{\dagger}.
\end{eqnarray}
By repeating this blocking procedure $\lceil \log_2(n-2)\rceil$ times
we can complete the decoupling, leaving a sum over
$O(D^{2\log_2 N}) = O(N^{2\log_2 D})$ terms involving the conjugation
of $H$ by local unitary operations.  Thus we can
decouple $P$ from $S$, leaving only the interaction on system $P$, using
a procedure of complexity $O(N^{2\log_2 D})$.  This interaction on
system $P$ can then be used to simulate an arbitary two-qudit interaction
on $P$, using the techniques described in the previous section.

\subsection{The case of localized two-qudit interactions}
\label{subsec:local}

%
%

The method just described assumes a general two-qudit Hamiltonian $H$.
Of course, the Hamiltonians occurring in Nature are usually much more
constrained.  In particular, it is very common for Hamiltonians to
have some sort of localized structure.  In this subsection we explain
how localized structure can be exploited to obtain more efficient
decoupling schemes than described above for the general case.  Note
that similar constructions in the context of NMR were
reported in~\cite{Jones99a,Leung00a}.

%
%

Suppose, for example, that $H$ contains only nearest-neighbour
interactions on a one-dimensional lattice.  This is obviously a
special case, but is a good illustration of the ideas used for more
general cases.  We number the qudits $1,2,\ldots,N$, and suppose that
$P$ contains qudits $1$ and $2$, while $S$ contains qudits $3$ through
$N$.  The case of general $P$ and $S$ follows using similar
techniques.  We can split the decoupling up into three steps.  In the
first step we eliminate all couplings between
$P$ and $S$, which can be achieved by
eliminating all couplings between qudits $2$ and $3$.  We call the resulting
Hamiltonian $H'$.  The second step is to eliminate all single-body
terms in $S$.  This can be done by simulating the Hamiltonian
\begin{eqnarray} H'' = \frac{1}{D^2}
\sum_U U_S H' U_S^{\dagger}.  \end{eqnarray} We
complete the decoupling by simulating \begin{eqnarray} 
H''' & = & \frac{1}{D^2} \sum_U
(I \otimes I \otimes U \otimes I \otimes U \otimes I
        \otimes \ldots) \times
        \nonumber \\
 & & H'' (I \otimes I \otimes U \otimes I \otimes U \otimes I \otimes
        \ldots)^{\dagger},
\end{eqnarray}
where the conjugation by $U$ is applied to qubits $3,5,7,\ldots$, which
we can easily see turns off all couplings acting between qudits in
$S$.

%
%

Thus, we see that for a nearest-neighbour Hamiltonian on a
one-dimensional lattice, the decoupling can be performed for {\em
constant} (with respect to $N$) cost in the simulation, as opposed to
the $O(N^{2 \log_2 D})$ cost incurred in the case when general
interaction terms appear in the Hamiltonian.

%
%

This result can easily be generalized.  Suppose
$S$ can be broken up into a partition $S_1,\ldots,S_m$ with the
property that qudits in one member of the partition $S_j$ only couple
to qudits outside $S_j$.  To decouple we do the
following.  For each element of the partition $S_j$ turn off all
couplings between $S_j$ and the remainder of the system by
simulating the Hamiltonian 
\begin{eqnarray} 
H_j = \sum_U U_{S_j} H_{j-1} U_{S_j}^{\dagger},
\end{eqnarray}
where $H_0 \equiv H$.  It is easy to see that the Hamiltonian $H_m$
contains no single-body terms from $S$, no couplings between $P$ and
$S$, and all couplings internal to $S$ have been eliminated.  The
total cost of the simulation scales as $(D^2)^m = D^{2m}$.  This cost
can be reduced even further by using a recursive procedure like that
described for the general two-qudit case, resulting in a scaling of
$O(D^{2 \log_2 m})$.

%
%

Many cases of interest can be described in the framework just
introduced.  For example, consider an $r$-dimensional cubic lattice of
qudits, with nearest-neighbour interactions. There is a natural
partitioning of this lattice into $2^r$ different sublattices, as follows.
First, fix a site in
the lattice, and then consider the cubic sublattice $S_1$ generated by
stepping $2$ lattice spacings in every direction.  We generate the
partition of sublattices $S_1,S_2,\ldots,S_{2^r}$ by translating $S_1$
one lattice spacing in various directions.  (We are ignoring boundary
conditions in this discussion; they are easily accommodated, or one can
imagine that the lattice has periodic boundary conditions).  Now
remove the qudits in $P$ from whichever elements $S_p$ and $S_q$ of
the partition they happened to fall into.  Notice that qudits in $S_j$
only ever couple {\em out} of $S_j$, since the interactions are
nearest-neighbour.  Thus the procedure described above makes it is
possible to decouple $P$ from $S$ using $O(D^{2r})$ operations.  More
generally, it is not difficult to use such constructions 
to efficiently decouple $P$ and $S$ for any
Hamiltonian containing only localized interactions.

\section{Discussion}
\label{sec:discussion}

%
%

We have shown that, given any two-qudit entangling Hamiltonian $H$ and
local unitaries we can simulate any other two-qudit Hamiltonian.  This
result was then applied to obtain universal gate constructions for
quantum computation.  Our results are of interest because they show
that such universal simulation is possible, in principle.  However,
the complexity of our construction limits the practicality of
potential implementations, and should encourage the search for more
practical methods.

%
%

There are two aspects to the analysis of efficiency for our
simulations.  The first is how they scale with the dimension $D$ of
the qudits in the system, and the second is how they scale with the
number $N$ of qudits present in the system.  The scaling with $N$ is
the critical factor, while the scaling with $D$ is not so important,
since for most physical systems of interest $D$ is a constant.  We
have shown that the scaling for simulation of one two-qudit
Hamiltonian with another is polynomial in $D$, and the scaling with
$N$ behaves as $O(N^{2\log_2 D})$.  Thus the total scaling is
$O(\mbox{poly}(D)N^{2\log_2 D})$, which is polynomial in both $N$ and
$D$.

%
%

Our results show that all two-body $N$-qudit entangling Hamiltonians
are qualitatively equivalent, given the ability to perform local
unitary operations.  Thus, in some sense the ability to entangle can
be regarded as a fundamental physical resource --- a type of ``dynamic
entanglement''\footnote{See footnote [37].} --- that can be utilized
to perform interesting processes.  It would be extremely interesting
to develop a detailed quantitative theory of such dynamic
entanglement.  Following the line of research we have pursued in this
paper, some potential questions one might attempt to answer in
developing such a theory of dynamic entanglement include:
\begin{itemize}

\item What is the optimal procedure for simulating one Hamiltonian
with another?  See~\cite{Wocjan02a,Bennett01a,Vidal01b} for preliminary
results in this direction.

\item Can an entangling Hamiltonian defined on a $D \times D'$ system,
  where $D \neq D'$, be used to perform universal simulation on those
  systems?  Note that this question has recently been settled in the
  affirmative~\cite{Wocjan02b,Bennett01c}, using methods rather
  different than that in our paper.

\item Our model assumes that the constituent systems are of finite
dimensionality $D$.  It would be interesting to determine whether
analogous results hold in infinite dimensions.

\item Are universal simulation results possible for non-unitary
processes?  For measurement processes?  Preliminary results in this
direction have been obtained in~\cite{Bacon01a,Lloyd02a}.

\item Can we weaken the condition that arbitrary local unitary
operations be allowed during the simulation procedure?  It would be
interesting, for example, if universal simulation could be performed in a
system where local unitaries are applied homogeneously across the
entire system.

\item Our model assumes that only a single Hamiltonian is being
applied at any given time, namely, either the entangling Hamiltonian
$H$, or a local Hamiltonian on a single qudit.  In practice, this is
not likely to be exactly the case.  What effect do imperfections have?

\item In the theory of entangled state transformation there is a
crucial distinction between ``single-shot'' manipulation of entangled
states, where just a single copy of the state is available, and
manipulations that are performed in the asymptotic limit where a large
number of copies of the state are available.  The results obtained in
the present paper are for single-shot Hamiltonian simulation; it would
be interesting to obtain results for the asymptotic case as well.

\end{itemize}

\section*{Acknowledgments} We thank Carl Caves for providing a copy of
Horn and Johnson at just the right time.  Especial thanks to Daniel
Gottesman for interesting discussions about the Pauli group, to Ben
Schumacher for many encouraging and motivating discussions, and to
Markus Grassl for pointing out a significant error in the first
version of this manuscript.  Thanks also to Ivan Deutsch, Tobias
Osborne, Damian Pope, and Rob Thew for helpful discussions.  AMC
thanks the Centre for Quantum Computer Technology at the University of
Queensland for its hospitality, and acknowledges the support of the
Fannie and John Hertz Foundation.  This work was supported in part by
the Department of Energy under cooperative research agreement
DF-FC02-94ER40818.

{\em Note:} After completion of this work we became aware that Wocjan,
Roetteler, Janzing and Beth have independently obtained some similar
results in~\cite{Wocjan02b}.

\appendix
\section{Normalizer operations for the $D$-dimensional Pauli group}
\label{app:Gottesman}

%
%

In this appendix we construct the unitary operations $U$ used to
perform the conjugation operations
Eqs.~(\ref{eq:Gott1})-(\ref{eq:Gott3}), which we reproduce here for
convenience,
\begin{eqnarray} 
X \rightarrow Z, & \,\, & Z \rightarrow X^{-1}; \label{eq:G1} \\ 
X \rightarrow XZ, & \,\, & Z \rightarrow Z; \label{eq:G2} \\ 
X \rightarrow X^a, & \,\, & Z \rightarrow 
Z^{a^{-1}}, \mbox{ provided } \gcd(a,D) = 1. \nonumber \\
 & &  \label{eq:G3}
\end{eqnarray} 
Our constructions are based on those of Gottesman~\cite{Gottesman99b},
however Gottesman's interest was mainly in the case of prime $D$
greater than $2$, and his constructions only apply for odd values of
$D$.  The following constructions apply for $D$ both odd and even.

%
%
The conjugation operation for Eq.~(\ref{eq:G1}) is just the
$D$-dimensional discrete Fourier transform, defined by
\begin{eqnarray}
U|j\rangle \equiv \sum_{k=0}^{D-1} \omega^{jk}|k\rangle
\end{eqnarray}
A straightforward calculation shows that $UXU^{\dagger} = Z$ and
$UZU^{\dagger} = Z^{-1}$, so Eq.~(\ref{eq:G1}) holds.

%
%
The definition of the conjugation operation for Eq.~(\ref{eq:G2})
depends on whether $D$ is odd or even.  When $D$ is odd we define
\begin{eqnarray}
U|j\rangle \equiv \omega^{j(j-1)/2}|j\rangle.
\end{eqnarray}
A straightforward calculation shows that $UXU^{\dagger} = XZ$ and
$UZU^{\dagger} = Z$, so Eq.~(\ref{eq:G2}) holds for odd $D$.  When $D$
is even we define
\begin{eqnarray}
U|j\rangle \equiv \omega^{j^2/2}|j\rangle,
\end{eqnarray}
and then check that $UXU^{\dagger} = \omega^{1/2}XZ$ and
$UZU^{\dagger} = Z$, so that Eq.~(\ref{eq:G2}) also holds for even
$D$.

%
%
Finally, the conjugation operation for Eq.~(\ref{eq:G3}) is defined by
\begin{eqnarray}
U|j\rangle \equiv |aj\rangle,
\end{eqnarray}
from which it follows that $UXU^{\dagger} = X^a$ and $UZU^{\dagger} =
Z^{a^{-1}}$, which completes the constructions needed to verify
  Eqs.~(\ref{eq:G1})-(\ref{eq:G3}).

\section{Second proof of the number theory lemma}
\label{app:number-theory}

%
%

In this appendix we provide an alternate proof of the number theory
lemma used in Subsection~\ref{subsec:detailed_proof} of the paper.
Recall the statement of the lemma:

\begin{quote} {\bf Lemma:} Suppose $\gcd(l,m) = 1$.  Then 
$j m =_D   k l$ if and only if there exists $n$ such that
\begin{eqnarray}
j =_D nl; \,\,\,\, k =_D nm.
\end{eqnarray}
\end{quote}

{\bf Proof:} By the PEG lemma there exists a normalizer operation $U$
such that $U X^l Z^m U^{\dagger} = Z^{\gcd(l,m)} = Z$, where the
equalities hold up to phase factors.  Note that $X^lZ^m$ commutes
with $X^j Z^k$, since $jm=_D kl$, so $UX^jZ^k U^{\dagger}$ must
commute with $Z = UX^lZ^mU^{\dagger}$.  It follows that
$UX^jZ^kU^{\dagger} = Z^n$, up to a phase factor, for some $n$, and
thus
\begin{eqnarray} X^jZ^k & = & U^{\dagger} Z^n U \\
        & = & (U^{\dagger} ZU)^n \\
        & = & (X^lZ^m)^n \\
        & = & X^{ln}Z^{mn},
\end{eqnarray}
where, again, the equalities hold up to unimportant phase factors.  It
follows that $j =_D ln$ and $k =_D mn$, as claimed. {\bf QED}


\end{document}